\documentclass[letter]{aa}  

\pdfoutput=1

\usepackage{graphicx}
\usepackage{longtable}
\usepackage{rotating}
\usepackage{epsfig}
\usepackage{natbib}
\usepackage{gensymb}
\usepackage{txfonts}

\usepackage[citecolor=blue, urlcolor=blue, linkcolor=blue, colorlinks=true]{hyperref}
\usepackage{cleveref}

\begin{document}

   \title{Searching for H$_{\alpha}$ emitting sources around MWC\,758
   \thanks{Based on observations obtained at Paranal Observatory under program 096.C-0267(A) and 96.C-0248(A).}}
   \subtitle{SPHERE/ZIMPOL high-contrast imaging}

   \author{N. Hu\'elamo\inst{1}
          \and
          G. Chauvin\inst{2,3}
          \and
          H. M. Schmid\inst{4}
          \and
          S. P. Quanz\inst{4}
          \and
          E. Whelan\inst{5}
          \and 
          J. Lillo-Box\inst{6}
          \and
          D. Barrado\inst{1}
          \and
          B. Montesinos\inst{1}
          \and
          J.M. Alcal\'a\inst{7}
          \and
          M. Benisty\inst{2,3}
          \and
          I. de Gregorio-Monsalvo\inst{6,8}
          \and
          I. Mendigut\'{\i}a\inst{1}
          \and
          H. Bouy\inst{9}
          \and
          B. Mer\'{\i}n\inst{10}
          \and
          J. de Boer\inst{11}
          \and
          A. Garufi\inst{12}
          \and
          E. Pantin\inst{13}
          }  
          \institute{Centro de Astrobiolog\'{\i}a (CSIC-INTA), Camino bajo del Castillo s/n, E-28692 Villanueva de la Ca\~nada, Madrid, Spain\\
              \email{nhuelamo@cab.inta-csic.es}
             \and
             Unidad Mixta Internacional Franco-Chilena de Astronom\'{i}a, CNRS/INSU UMI 3386 and Departamento de Astronom\'{i}a, Universidad de Chile, Casilla 36-D, Santiago, Chile
             \and
             Univ. Grenoble Alpes, CNRS, IPAG, F-38000 Grenoble, France
             \and
             ETH Zurich, Institute of Particle Physics and Astrophysics, Wolfgang-Pauli-Strasse 27, 8093 Zurich, Switzerland
             \and
             Maynooth University Department of Experimental Physics, National University of Ireland, Maynooth Co. Kildare, Ireland
              \and
             European Southern Observatory, Alonso de Cordova 3107, Casilla 19, Vitacura, Santiago, Chile
             \and
             INAF-Osservatorio Astronomico di Capodimonte, via Moiariello 16, 80131 Napoli, Italy
                          \and 
             Joint ALMA Observatory, Alonso de C\'ordova 3107, Vitacura, Santiago, Chile
             \and
             Laboratoire d'Astrophysique de Bordeaux, Univ. Bordeaux, CNRS, B18N, all\'ee Geoffroy Saint-Hilaire, 33615 Pessac, France
             \and
             ESAC Science Data Centre, ESA-ESAC, Spain
             \and
             Leiden Observatory, Leiden University, PO Box 9513, 2300 RA Leiden, The Netherlands
             \and
             Universidad Aut\'onoma de Madrid, Dpto. F\'{\i}sica Te\'orica, Madrid, Spain
              \and
             Laboratoire AIM, CEA/DRF, CNRS, Universit\'e Paris Diderot, IRFU/DAS, F-91191 Gif sur Yvette, France
             }   
           \date{Received ; accepted }

 \abstract
   {MWC\,758 is a young star surrounded by a transitional disk. The disk shows an inner cavity and spiral arms that could be caused by the presence of  protoplanets. Recently, a protoplanet candidate has been detected around MWC\,758 through high-resolution $L'$-band observations. The candidate is located inside the disk cavity at a separation of $\sim$111\,mas from the central star, and at an average position angle of $\sim$165.5$\degree$. 
   }
   {We aim at detecting accreting protoplanet candidates within the disk of MWC\,758 through spectral angular differential imaging observations in the optical regime. In particular, we explore the emission at the position of the detected planet candidate.}
   {We have performed simultaneous adaptive optics observations in the H$_{\alpha}$ line and the adjacent continuum 
   using SPHERE/ZIMPOL at the Very Large Telescope (VLT).}
   {The data analysis does not reveal any H$_{\alpha}$ signal around the target. 
   The derived contrast curve in the B\_Ha filter allows us to derive a 
   5$\sigma$ upper limit of $\sim$7.6\,mag at 111\,mas, the separation of the previously detected planet candidate. 
   This contrast  translates into a H$_{\alpha}$ line luminosity of  $L_{\rm H_{\alpha}}\lesssim$ 5$\times$10$^{-5}$ $L_{\odot}$ at 111\,mas.  
   Assuming that $L_{\rm H_{\alpha}}$ scales  with $L_{\rm acc}$ as in Classical T Tauri stars as a first approximation, we can estimate an accretion luminosity  of $L_{acc} <$3.7$\times$10$^{-4}\,L_{\odot}$ for the protoplanet candidate. 
   For the predicted mass range of MWC\,758b, 0.5-5\,$M_{\rm Jup}$, this implies accretion rates smaller than  $\dot M<$\,3.4$\times ($10$^{-8}$-10$^{-9})\,M_{\odot}/yr$, for an average planet radius of 1.1\,$R_{\rm Jup}$. Therefore, our estimates are consistent with the predictions of accreting circumplanetary accretion models for $R_{\rm in} = 1 R_{\rm Jup}$. The ZIMPOL line luminosity is consistent with the H$_{\alpha}$ upper limit predicted by these models for truncation radii  $\lesssim$3.2\,R$_{\rm Jup}$.} 
   {The non-detection of any H$_{\alpha}$ emitting source in the ZIMPOL images does not allow us to unveil the nature of the $L'$  detected source. Either it is a protoplanet candidate or a disk asymmetry. }
   
  \keywords{stars: pre-main sequence  --  stars: planetary systems -- stars: individual: MWC\,758 -- accretion, accretion disks
               }
\maketitle


\section{Introduction}

The study of transitional disks is an important tool to understand planet formation and evolution. 
High-angular-resolution observations of these disks at different wavelengths have revealed structures including spiral arms, warps, gaps and radial streams that might be related to on-going planet formation 
\citep[e.g.][]{Andrews2011,Mayama2012,Grady2013,Boccaletti2013,Pinilla2015,Perez2016,Mendigutia2017, Pohl2017}. 
Therefore, a great effort has been made to detect protoplanets 
still embedded in these disks 
\citep[e.g.][]{Huelamo2011,Kraus2012,Quanz2013,Biller2014,Close2014,Whelan2015,Sallum2015}, 
since they can shed light on the planet formation mechanisms, the required physical conditions, and 
the physics of gas accretion to form the atmospheres of giant planets.

In this context, MWC\,758 (HD\,36112, HIP25793) is an interesting target to study planet formation. 
The object is a young \citep[3$\pm$2\,Myr,][]{Meeus2012} Herbig Ae star. 
While the  {\it Hipparcos} parallax provided a distance of 279$^{+94}_{-58}$\,pc \citep{vanLeeuwen2007}, the $GAIA$ TGAS Catalog  reports a parallax of 6.63$\pm$0.38 mas, resulting in a distance of 151$^{+10}_{-8}$\,pc  \citep{GaiaDR12016,Gaia2016}.

MWC\,758 is surrounded by a transitional disk spatially resolved at infrared (IR) and sub-milimeter (sub-mm) wavelengths \citep{Chapillon2008,Isella2010, Andrews2011,Grady2013, Marino2015}. The disk inclination is 21$\pm$2$\degree$, and its shows several structures and a marginally resolved cavity.
 \citet{Benisty2015} presented SPHERE/IRDIS IR polarimetric observations of the source in 
the Y-band (1.04\,$\mu$m). The unprecedented angular resolution of these observations allowed them to spatially resolve several non-axisymmetric features down to 26\,au ($\sim$14\,au with the new $GAIA$ distance), and a no fully depleted cavity.  They also resolved the two spiral arms previously reported by \citet{Grady2013} through HiCIAO observations. They presented a model that concludes that 
the presence of planets inside the cavity cannot reproduce the large opening angle of the spirals. On the other hand, 
models with an outer companion could work better to explain these disk structures \citep[see e.g.][]{Dong2015}.
More recently, \citet{Boehler2017} presented new sub-mm ALMA data showing a large dust cavity of $\sim$40\,au in radius, evidence of a warped inner disk, and the presence of two dust clumps in the outer regions of the disk. To explain the whole disk structure, they propose the presence of two giant planets,  one in the inner regions responsible for carving the cavity, and an outer one responsible for the spirals.

Interestingly,  \citet{Reggiani2017} reported the presence of a point-like source inside the sub-mm cavity, at a separation of $\sim$111$\pm$4\,mas 
from the central star. The object was detected in the L-band at two different epochs, with a contrast of
$\Delta L$=7.0$\pm$0.3\,mag with respect to the primary. As the authors explain, this emission could be caused by an embedded protoplanet, although they cannot exclude that  it is associated with an asymmetric disk feature. The comparison of their observations  with circumplanetary disk accretion models \citep{Zhu2015} are consistent with a 0.5-5\,$M_{\rm Jup}$ planet accreting at a rate of 10$^{-7}$-10$^{-9}$\,$M_{\odot}/yr$.

We have explored the possibility of detecting protoplanet candidates using the H$_{\alpha}$ emission line as an accretion tracer
\citep[see e.g.][]{Close2014,Sallum2015}. In this letter, we present  spectral angular differential imaging observations of MWC\,758 obtained with SPHERE/ZIMPOL at the Very Large Telescope (VLT).  We used  two filters, one centered at the H$_{\alpha}$ line,  and the other one at the adjacent continuum.  We 
complemented the ZIMPOL data with optical spectroscopy using the  CAFE spectrograph at the 2.2m telescope in the Calar Alto observatory. MWC\,758 was part of a program to detect young accreting protoplanets within the circumstellar disks of young stars.


\section{Observations and data reduction}

\subsection{VLT/ZIMPOL}

The SPHERE Open Time observations (096.C-0267.A) were obtained on December 30, 2015.  The ZIMPOL instrument  of  SPHERE  \citep{beuzit2008,thalmann2008} was used in spectral and angular differential imaging modes \citep{marois2006,racine1999}. In addition to the pupil stabilized mode, ZIMPOL simultaneously imaged MWC\,758 in two different filters: B\_Ha ($\lambda_{c} = 655.6$\,nm and $ \Delta \lambda = 5.5$\,nm) and Cnt\_Ha ($\lambda_{c} = 644.9$\,nm and $ \Delta \lambda = 4.1$\,nm). We obtained 190 individual exposures of 60 seconds each, resulting in a total exposure time of $\sim$3 hours on-source (from 02:20\,UT to 05:24\,UT).  The conditions were photometric until 03:39 (UT) when they turned into clear.

\begin{figure}[t!]
 \includegraphics[width=9.0cm]{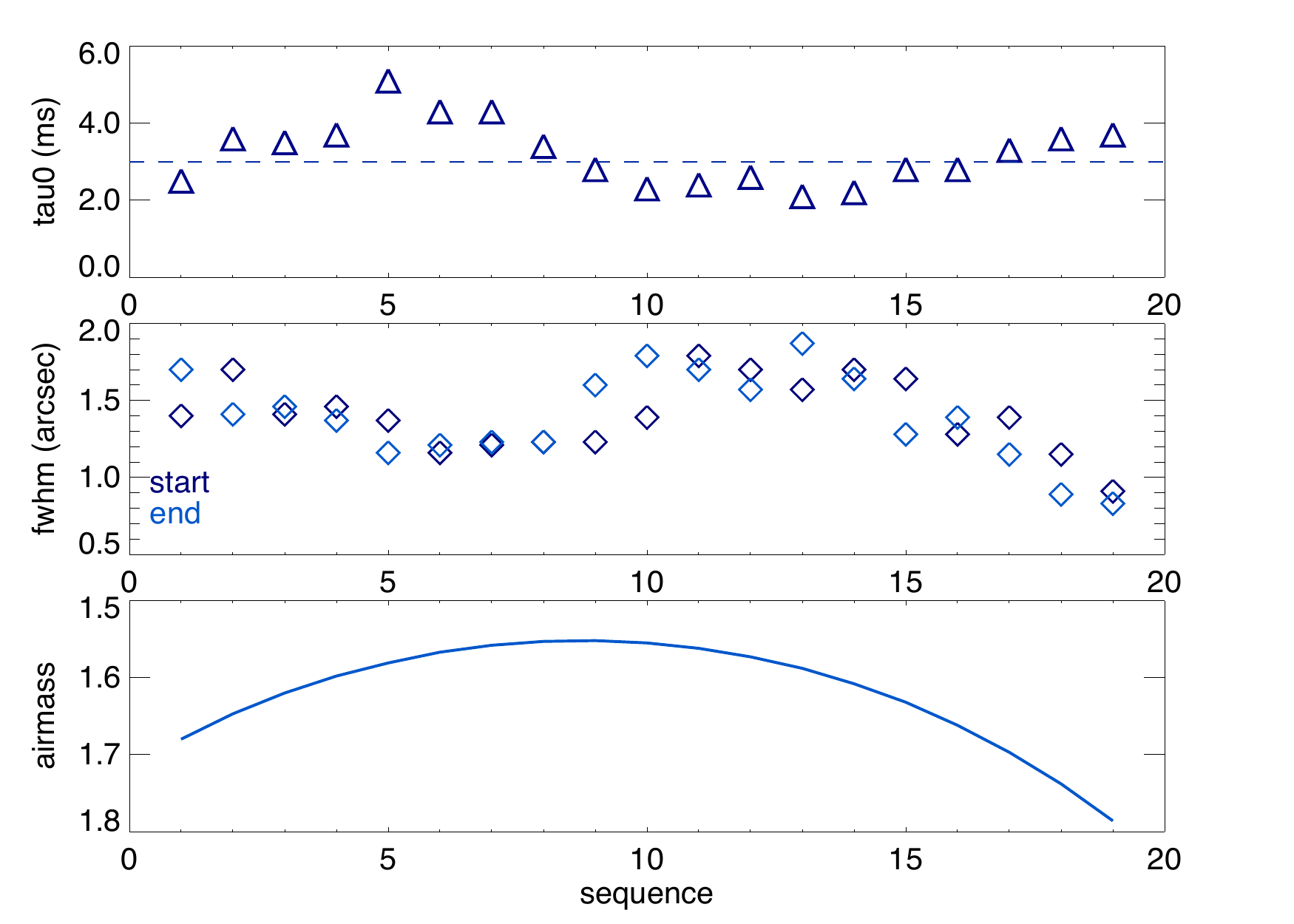}
 \caption{Average atmospheric conditions during the ZIMPOL observations.The x-axis shows the average conditions in 
 cycles of 10~mins. Therefore, we have represented 19 cycles of 10~mins each. From top to bottom, we have plotted the coherence time ($\tau_0$), the average FWHM as seen by the guide probe at the beginning and end of the cycles, and the airmass.}\label{conditions}
 \end{figure}

\begin{figure}[t!]
 \includegraphics[width=9cm]{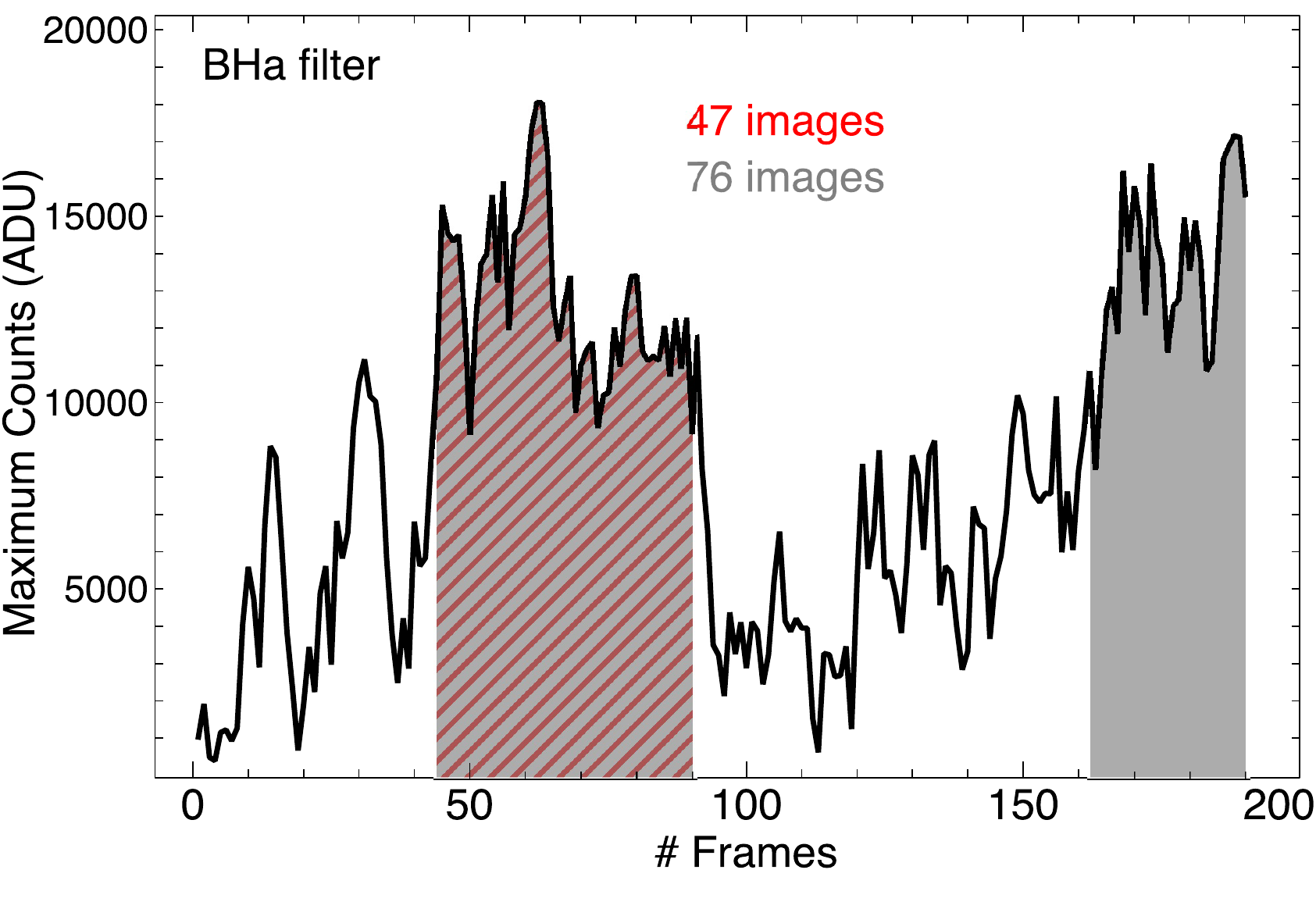}
 \caption{The maximum counts (in ADU) measured in the 190 individual frames obtained in the B\_Ha filter. 
The gray shaded areas correspond to the 76 selected images (see text), while the  
hatched red area corresponds to the best 47 individual exposures.}\label{FluxStar}
 \end{figure}

 The observing conditions (monitored by the seeing and coherence time) were variable during the complete observing sequence. 
 The seeing varied between 0\farcs7 and 2\farcs0, and the coherence time between 5\,ms and 2\,ms (see Fig~\ref{conditions}) affecting the final Adaptive Optics (AO) correction of the ZIMPOL images. Figure~\ref{FluxStar} 
 shows the maximum counts registered on the target (good proxy for the Strehl variation in 
 photometric/clear conditions) throughout the 190 individual images in the B\_Ha filter (the same variations are observed in Cnt\_Ha).  
The mean full-width at half-maximum (FWHM)  measured in the individual images is $\sim$9 pixels, that is,  $\sim$32\,mas considering the pixel scale of 3.6\,mas/pixel of the ZIMPOL detector.
 
  The data reduction was performed using a customary pipeline developed by our team. As a first step, the data were bias-subtracted, flat-fielded, and bad-pixel corrected. Subsequently, the individual images were recentered using a simple Moffat function to estimate the centroid position as no coronograph was used. For the point-spread function (PSF) subtraction, the individual filter datasets (B\_Ha and Cnt\_Ha) were considered separately to apply a standard Angular Differential Imaging (ADI) processing technique using classical and smart-ADI (cADI and sADI; \citealt{,Lafreniere2007,Chauvin2012}) and principal component analysis (PCA) \citep{Soummer2012}. For the B\_Ha-Cnt\_Ha spectral and angular differential imaging processing, the individual Cnt\_Ha images were first spatially rescaled to the B\_Ha filter resolution, then flux-normalized considering the total flux ratio between B\_Ha and Cnt\_Ha within an aperture of $r=10$ pixels, before finally applying the subtraction B\_Ha-Cnt\_Ha frame per frame to exploit the simultaneity of both observations. ADI processing in cADI, sADI and PCA was then applied to the resulting differential datacube.  

 \begin{figure}
 \includegraphics[scale=0.45]{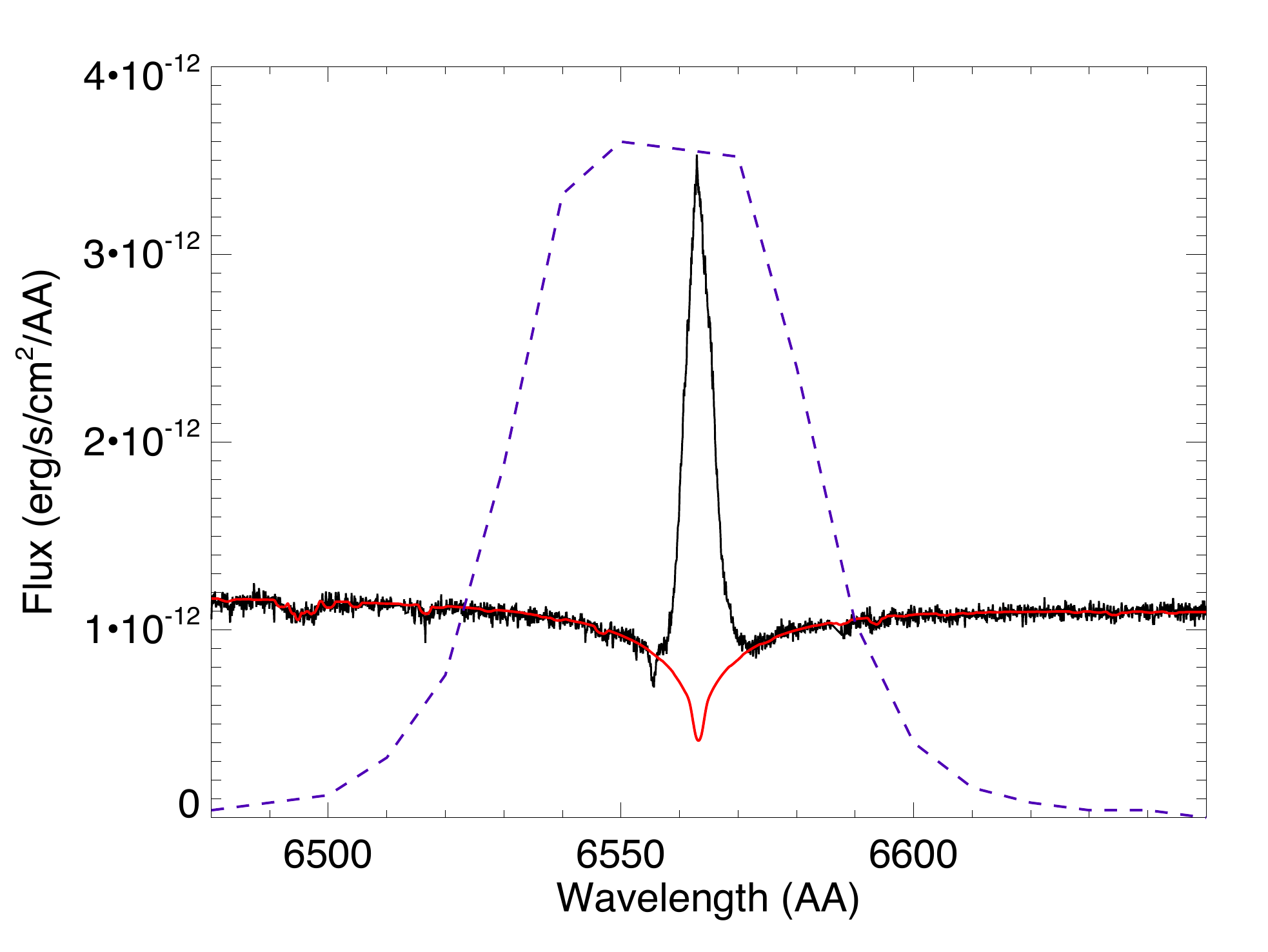}
 \caption{The observed CAFE/CAHA spectrum around the H$_{\alpha}$ line (in black). The red line represents the model used to fit the stellar atmosphere (see text). The dashed blue line represents the ZIMPOL B\_Ha filter transmission curve.}\label{specCAFE}
 \end{figure}

The ADI detection limits were then estimated from a standard pixel-to-pixel noise map of each filter within a box of 5$\times$5 pixels sliding from the star to the limit of the ZIMPOL field of view. The contrast curves at $5\sigma$ were then
obtained using the pixel-to-pixel noise map divided by the ADI flux loss, and corrected from small number statistics following the prescription of \citet{Mawet2014} adapted to our $5\sigma$ confidence level at small angles with ZIMPOL. The same method was applied for the differential imaging dataset to illustrate the gain in speckle subtraction in the inner region below 200\,mas. 

Finally, and to consider the impact of variable atmospheric conditions, we analyzed three different datasets
(see Figure~\ref{FluxStar}): first, we worked with the whole observing sequence (190 images) which covers a field rotation of $\sim$55 degrees; second,  we selected the best AO-corrected images through the whole exposure (images 44 to 90, and 162 to 190, a total of 76 frames, covering 15\,deg and 7\,deg of field rotation, respectively) and, third, we kept the 47 images with the best AO correction in the first half of the observing sequence (images 44 to 90, $\sim$15 deg of field rotation). We finally worked with the latest dataset (47 individual images) to achieve the best detection performances at close separations. The average Strehl ratio of this dataset is $\sim$10\%.
 
 \subsection{CAHA/CAFE spectroscopy}
 
 To complement the SPHERE/ZIMPOL observations, we performed high-resolution spectroscopy of MWC\,758 with the CAFE spectrograph attached to the 2.2m telescope in the Calar Alto Observatory \citep{Aceituno2013}. CAFE is equipped with a 2\farcs4 diameter optical fibre, and provides 
 high-resolution spectra (R$\sim$63\,000) over the 3900-9600\AA~spectral range. 
 
 We obtained a spectrum on the night of December 24 2015 under photometric conditions and with an average seeing of 0\farcs7.   The total exposure time was 600 seconds. The spectrum was reduced with a dedicated pipeline developed by  the  observatory and explained  in  \cite{Aceituno2013}.  The processing was as follows: the data were bias subtracted and flat-field corrected.  We  used  thorium-argon (ThAr) exposures obtained after the science spectrum to wavelength-calibrate the data. The spectrum was flux-calibrated using the standard star HD19445, observed one hour before the
science target with an airmass difference of $\sim$0.2. We compared the calibrated spectrum with published photometry \citep{Beskrovnaya1999}, obtaining that the spectrum is $\sim$1.2 times brighter (0.2\,mag) in the R-band.  
Therefore, we estimate a precision in the flux calibration of $\sim$20\% in this band.

Figure~\ref{specCAFE} shows a portion of the calibrated CAFE spectrum containing the H$_{\alpha}$ line.
We measure a line flux of 1.3$\pm$0.3$\times$10$^{-11}$ erg/s/cm$^2$.
To correct for the stellar photospheric absorption, we first computed synthetic spectra using the codes 
{\sc atlas} and {\sc synthe} \citep{Kurucz1993} fed with the models describing the
stratification of the stellar atmospheres \citep{Castelli2003}. Solar abundances were assumed.
The synthetic spectrum was computed assuming $T_{\rm eff}$=7750 K and $\log g_*$=4.0 as the starting point,
following the result from \citet{Beskrovnaya1999}.
A value of $v \sin i$=50 km/s was adopted since it reproduces well the width of the 
photospheric line profiles. The model is displayed as a red line in Figure~\ref{specCAFE}.
The estimated H$_{\alpha}$ line flux, corrected from the photospheric absorption, is 
1.6$\pm0.3\times$10$^{-11}$ erg/s/cm$^2$. Taking into account the reported visual extinction of $A_{\rm V}$=0.16\,mag \citep{Meeus2012}, we can correct the flux adopting a value of $A_{\rm R}\sim0.12$\,mag \citep{Rieke85}, and estimate a line luminosity of $\sim$0.013\,$L_{\odot}$ for a distance of 151\,pc.

Assuming that the Herbig Be/Ae relationship between H$_{\alpha}$ luminosity and
accretion luminosity derived by   \citet{Fairlamb2017} holds, we estimate 
$\log\,(L_{\rm acc}/L_{\odot}) = (2.09\pm0.06) + (1.00\pm0.05) \times \log(L_{\rm H_{\alpha}}/L_{\odot})$.
We derive $L_{\rm acc}=1.6\pm0.4\,L_{\odot}$.  Following \citet{Mendigutia2011}, we estimate a mass accretion 
rate of $\sim(7\pm2)\times$10$^{-8}$\,$M_{\odot}/yr$,  for a stellar mass of 1.4$\pm$0.3\,$M_{\odot}$ and a stellar radius of
2\,$R_{\odot}$ \citep{Boehler2017}, already scaled to the {\it GAIA} distance.

Finally, we note that although the CAFE spectrum is not simultaneous to the SPHERE observations, it allows us to 
characterize the primary star and to double-check the calibration of our ZIMPOL data with close in-time observations.

 \section{Results and discussion}

\begin{figure*}[ht!]
 \includegraphics[scale=0.92]{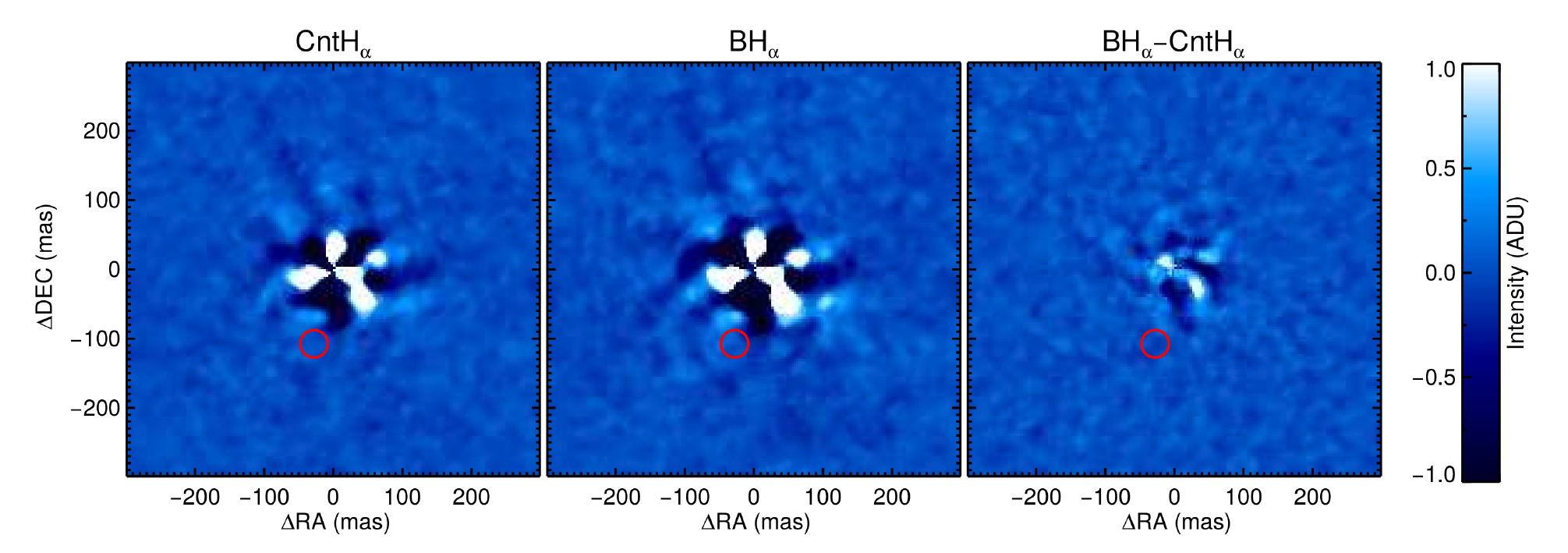}
 \caption{VLT/ZIMPOL ADI reduced images of MWC\,758 processed with PCA (5 modes) in the Cnt\_Ha (left) and B\_Ha (center) filters. The ASDI image (B\_Ha-Cnt\_Ha) processed with PCA (5 modes) is displayed in the right panel. North is up and East to the left. The red circle shows the position at a separation of 111\,mas, and a PA of 165.5\degree, where the
 companion candidate was detected by \citet{Reggiani2017} in the $L'$-band.}\label{asdi}
 \end{figure*}
 

 \begin{figure}[t!]
 \begin{center}
 \includegraphics[scale=0.48]{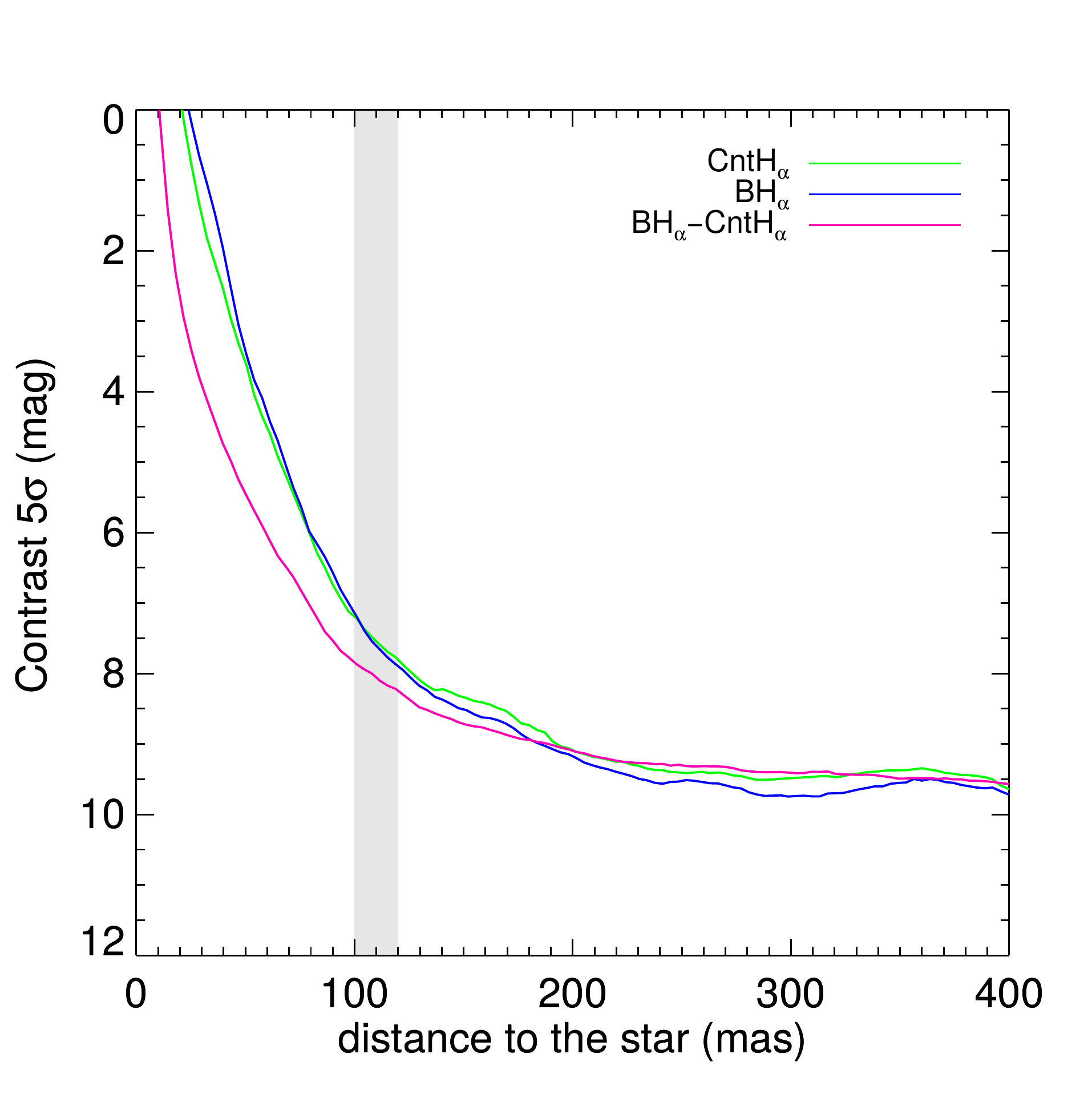} 
 \caption{VLT/ZIMPOL $5 \sigma$ contrast curves obtained in Cnt\_Ha and B\_Ha in ADI using PCA, 
 and in B\_Ha-Cnt\_Ha in ASDI using PCA. The gray area shows the position of the companion candidate 
 detected at $L'$ by \cite{Reggiani2017}. The curves are derived using the best 47 images.}
 \label{contrast}
 \end{center}
 \end{figure}

 The SPHERE/ZIMPOL combined images of MWC\,758 in the B\_Ha and the Cnt\_Ha filters, together with the differential 
 B\_Ha--Cnt\_Ha  image, are displayed in Figure~\ref{asdi}.  We show the results of combining the 47 best images. We have not detected any point-like source in any of the analyzed datasets. The  5$\sigma$ contrast achieved in the two individual filters, and in the differential sequence,  is displayed in Figure \ref{contrast}. We reach $\sim$7.6\,mag in the B\_Ha and the Cnt\_Ha filters  at a separation of 111\,mas, that is, where the protoplanet candidate by \citet{Reggiani2017} is detected.

The fact that we do not detect any companion in the individual filters and differential images does not allow us a direct
estimation of its $H_{\alpha}$ flux. However, since the 47 images were observed under photometric conditions, 
we can convert the  B\_Ha contrast curve into fluxes by calibrating the counts ($cts$) from the primary star. To this aim, we first performed aperture photometry of the central star in ten individual images (out of the 47), using a large aperture radius (320 pixel radius, $\sim$1\,arcsec). We obtain a total  of $\sim$6.5($\pm$0.6)$\times$10$^6$ $cts$. The uncertainty includes a relative error of 10\% associated to the channel splitting, and the background error estimation. We converted the integrated counts into fluxes using the zero-points (ZPs) derived by \citet{Schmid2017}, 
and their Eq. (4):  
\begin{equation}
{\tiny
f = (cts/s) \cdot 10^{0.4(a_m \times k_1 + m_{mode})} \cdot c_{zp}^{l}
}
\end{equation}\label{ecuacion1}

Assuming that all the emission in the B\_Ha filter comes from the H$_\alpha$ line, we can convert the counts per second 
adopting a zero-point ($c_{zp}^{l}$) of 7.2($\pm$0.4)$\times$10$^{-16}$ erg/(cm$^2$ ct), and an $m_{mode}$ value of $-$0.23\,mag  to take into account the usage of the $R$-band dichroic (Schmid, private communication). For an average airmass ($a_m$) of 1.57 (for the 47 images), and a typical $R$-band atmospheric extinction value of $k_{1}\sim$0.08 \citep{Patat2011}, we obtained a B\_Ha flux of $\sim7.1(\pm$0.8)$\times$10$^{-11}$ erg/s/cm$^2$ for the primary star.  
We note that the assumption that all the counts in the B\_Ha filter come from the line, results in a line flux overestimation of a factor of $\sim$5 when compared with the line flux measured directly in the spectrum. 

We used the total calibrated flux in B\_Ha, together with the contrast curve, to derive an upper limit 
(5-$\sigma$) to the B\_Ha  flux at different separations form the central object. In particular, we estimated a flux $<6.2\times$10$^{-14}$  erg/s/cm$^2$ at 111\,mas, the separation of the planet candidate.  If we assume that all the flux comes from the line, and an extinction equal to that of the primary star ($A_{\rm R}$=0.12\,mag), we estimate a line luminosity of $L_{\rm H_{\alpha}} <$4.9$\times$10$^{-5}\,L_{\odot}$ for a distance of 151\,pc. 
When compared with the other two H$_{\alpha}$ sources detected in transitional disks, we see that our ZIMPOL upper limit is of 
the same order of the emission of the protoplanet candidate LkCa15~b \citep[6$\times10^{-5}L_{\odot}$,][]{Sallum2015}, and one order of magnitude fainter than the very low-mass star ($\sim$0.2\,M$_{\odot}$) HD142527~B \citep[5$\times10^{-4}L_{\odot}$,][]{Close2014}.
We note that the smallest and lightest accreting companions to stars 
in the brown dwarf/planet boundary (e.g., GQ Lup~b, FW~Tau~b, GSC 06214--0210~b, DH~Tau~b) show  H$_{\alpha}$ luminosity values 
in the range $L_{\rm H_{\alpha}}\sim2\times10^{-5}$-- 6.5$\times 10^{-7} L_{\odot}$  \citep[see e.g.][]{Bowler2014,Zhou2014}. 

As explained in \cite{Reggiani2017}, the mass of the companion candidate ($M_p$) should be below 5.5\,$M_{\rm Jup}$  
to allow for small dust replenishment within the disk. The comparison of their $L'$-band luminosity with circumplanetary accretion models 
\citep{Zhu2015} is consistent with $M_p\dot M\sim2(\pm1)-6.0(\pm1.5)\times10^{-5}\,M_{\rm Jup}^2/yr$ for disk inner radii of $R_{in}=1-4\,R_{\rm Jup}$, respectively (see their Fig. 6). Given the mass constraints, their data is compatible with a 0.5-5\,$M_{\rm Jup}$ protoplanet with an accretion rate of 10$^{-7}$-10$^{-9}\,M_{\odot}/yr$.

As a first approximation, and following previous works \citep{Close2014,Sallum2015}, we can estimate the accretion luminosity 
($L_{\rm acc}$) of MWC\,758~b assuming that it scales with $L_{\rm H_{\alpha}}$ as in Classical T Tauri stars (CTTSs). In this case, 
$L_{\rm acc} = 10^{[(2.99\pm0.16) + (1.49\pm0.05)*\log (L_{\rm H_{\alpha}})]}$  \citep{Rigliaco2012}, and we derive an upper limit of 
$L_{\rm acc} < 3.7\times10^{-4}\,L_{\odot}$ at 111\,mas.  
Using the  $L_{\rm acc}$ upper limit, and  following  \citet{Gullbring1998}, we estimate 
$M_p \dot M < 1.8\times10^{-5}\,M_{\rm Jup}^{2}/yr$, assuming a planet radius of 1.1\,$R_{\rm Jup}$ 
\citep[the average value for 0.5-5$M_{\rm Jup}$ planets according to 3\,Myr COND models,][]{Baraffe2003}.  
This upper limit is consistent with the lowest values of $M_p \dot M$ derived for $R_{\rm in}$=1\,$R_{\rm Jup}$
by \citet{Reggiani2017}.  If we assume planet masses of 0.5-5\,$M_{\rm Jup}$ and a planet radius of 1.1\,$R_{\rm Jup}$, 
we estimate accretion rates of $\dot M < 3.4\times(10^{-8}-10^{-9})\,M_{\odot}/yr$ for MWC\,758~b. 

We note, however, that low-mass brown dwarfs and planetary mass objects seem not to follow the  $L_{\rm H_{\alpha}}$-- $L_{\rm acc}$ relation derived for CTTSs: as an example, \citet{Zhou2014} estimated the accretion rates of three very low-mass substellar sources through their
UV/optical excesses, and found values more than one order of magnitude higher than the values expected from extrapolating the CTTSs relationship. 
Interestingly, they also show that, in contrast with young TTSs, very low-mass accreting objects might emit a larger fraction of their accretion  luminosity in the H$_{\alpha}$ line. 

An alternative way to study the H$_{\alpha}$ emission of protoplanets is to compare their estimated luminosity with predictions from accreting circumplanetary models \citep{Zhu2015}. In this latter work, the author studied the expected H$_{\alpha}$ emission from protoplanets due to magnetospheric accretion (see their Eq. 22), obtaining upper limits to their $L_{\rm H_{\alpha}}$ luminosity three orders of magnitude lower than the typical values derived for CTTs  \citep[$\sim$5$\times10^{-3} L_{\odot}$,][]{Rigliaco2012}. The line luminosity depends on the truncation radius of the inner disk, and the infall velocity of the accreted material. According to their Eq. 22, our ZIMPOL data ($L_{\rm H_{\alpha}} \lesssim$  5$\times10^{-5}\,L_{\odot}$) is consistent with their predicted luminosity for truncation radii smaller
than $\sim$3.2\,R$_{\rm Jup}$.

The fact that we do not detect any source in the B\_Ha filter might suggest that the MWC\,758 protoplanet candidate 
is accreting at a low rate and we are not sensitive to its H$\alpha$ emission. Another possibility is that the object
is undergoing episodic accretion and we have observed it in a quiescent state. 
Although the disk inclination is closer to face-on, the sub-mm cavity is filled with small grains \citep{Benisty2015}, so the object might also be too extincted and therefore undetectable in the optical regime. 
Finally, we cannot exclude the possibility that the $L'$ infrared emission is related to a disk asymmetry and not with a protoplanet, as proposed by \citet{Reggiani2017}. In fact, it is known that ADI techniques can highlight disk features in such a way that they can  look like 
planets, especially for low-inclination systems \citep[see e.g.][]{Ligi2018}. 
Follow-up $L'$ observations, together with deeper $K$-band data, will help to understand the true nature of the infrared detection.

 \begin{acknowledgements}
 
We are grateful to the referee, C. Grady, whose comments helped to improve this manuscript.
This research has been funded by Spanish grants ESP2015-65712-C5-1-R and ESP2017-87676-C5-1-R.
NH is indebted to the Paranal staff that gave their support during the observations, and to the CAHA staff.
Part of this work has been carried out within the framework of the National Centre for Competence in Research PlanetS
supported by the Swiss National Science Foundation. S.P.Q. and
H.M.S. acknowledge the financial support of the SNSF.
JMA acknowledge financial support from the project PRIN-INAF 2016
The Cradle of Life - GENESIS-SKA (General Conditions in Early Planetary
Systems for the rise of life with SKA). I.M. acknowledges the financial support of the Government of Comunidad
Aut\'onoma de Madrid (Spain) through a "Talento'' Fellowship
(2016-T1/TIC-1890). Based on observations collected at the Centro Astron\'omico Hispano Alem\'an (CAHA) at Calar Alto, operated jointly by the Max-Planck Institut f\"ur Astronomie and the Instituto de Astrof\'{\i}sica de Andaluc\'{\i}a (CSIC)'.
This work has made use of data from the European Space Agency (ESA)
mission {\it Gaia} (\url{https://www.cosmos.esa.int/gaia}), processed by
the {\it Gaia} Data Processing and Analysis Consortium (DPAC,
\url{https://www.cosmos.esa.int/web/gaia/dpac/consortium}). Funding
for the DPAC has been provided by national institutions, in particular
the institutions participating in the {\it Gaia} Multilateral Agreement. 
\end{acknowledgements}

 \bibliographystyle{aa}
 \bibliography{mwc758}

\begin{thebibliography}{50}
\expandafter\ifx\csname natexlab\endcsname\relax\def\natexlab#1{#1}\fi

\bibitem[{{Aceituno} {et~al.}(2013){Aceituno}, {S{\'a}nchez}, {Grupp}, {Lillo},
  {Hern{\'a}n-Obispo}, {Benitez}, {Montoya}, {Thiele}, {Pedraz}, {Barrado},
  {Dreizler}, \& {Bean}}]{Aceituno2013}
{Aceituno}, J., {S{\'a}nchez}, S.~F., {Grupp}, F., {et~al.} 2013, \aap, 552,
  A31

\bibitem[{{Andrews} {et~al.}(2011){Andrews}, {Wilner}, {Espaillat}, {Hughes},
  {Dullemond}, {McClure}, {Qi}, \& {Brown}}]{Andrews2011}
{Andrews}, S.~M., {Wilner}, D.~J., {Espaillat}, C., {et~al.} 2011, \apj, 732,
  42

\bibitem[{{Baraffe} {et~al.}(2003){Baraffe}, {Chabrier}, {Barman}, {Allard}, \&
  {Hauschildt}}]{Baraffe2003}
{Baraffe}, I., {Chabrier}, G., {Barman}, T.~S., {Allard}, F., \& {Hauschildt},
  P.~H. 2003, \aap, 402, 701

\bibitem[{{Benisty} {et~al.}(2015){Benisty}, {Juhasz}, {Boccaletti},
  {Avenhaus}, {Milli}, {Thalmann}, {Dominik}, {Pinilla}, {Buenzli}, {Pohl},
  {Beuzit}, {Birnstiel}, {de Boer}, {Bonnefoy}, {Chauvin}, {Christiaens},
  {Garufi}, {Grady}, {Henning}, {Huelamo}, {Isella}, {Langlois}, {M{\'e}nard},
  {Mouillet}, {Olofsson}, {Pantin}, {Pinte}, \& {Pueyo}}]{Benisty2015}
{Benisty}, M., {Juhasz}, A., {Boccaletti}, A., {et~al.} 2015, \aap, 578, L6

\bibitem[{{Beskrovnaya} {et~al.}(1999){Beskrovnaya}, {Pogodin},
  {Miroshnichenko}, {Th{\'e}}, {Savanov}, {Shakhovskoy}, {Rostopchina},
  {Kozlova}, \& {Kuratov}}]{Beskrovnaya1999}
{Beskrovnaya}, N.~G., {Pogodin}, M.~A., {Miroshnichenko}, A.~S., {et~al.} 1999,
  \aap, 343, 163

\bibitem[{{Beuzit} {et~al.}(2008){Beuzit}, {Feldt}, {Dohlen}, {Mouillet},
  {Puget}, {Wildi}, {Abe}, {Antichi}, {Baruffolo}, {Baudoz}, {Boccaletti},
  {Carbillet}, {Charton}, {Claudi}, {Downing}, {Fabron}, {Feautrier},
  {Fedrigo}, {Fusco}, {Gach}, {Gratton}, {Henning}, {Hubin}, {Joos}, {Kasper},
  {Langlois}, {Lenzen}, {Moutou}, {Pavlov}, {Petit}, {Pragt}, {Rabou}, {Rigal},
  {Roelfsema}, {Rousset}, {Saisse}, {Schmid}, {Stadler}, {Thalmann}, {Turatto},
  {Udry}, {Vakili}, \& {Waters}}]{beuzit2008}
{Beuzit}, J.-L., {Feldt}, M., {Dohlen}, K., {et~al.} 2008, in \procspie, Vol.
  7014, Ground-based and Airborne Instrumentation for Astronomy II, 701418

\bibitem[{{Biller} {et~al.}(2014){Biller}, {Males}, {Rodigas}, {Morzinski},
  {Close}, {Juh{\'a}sz}, {Follette}, {Lacour}, {Benisty}, {Sicilia-Aguilar},
  {Hinz}, {Weinberger}, {Henning}, {Pott}, {Bonnefoy}, \&
  {K{\"o}hler}}]{Biller2014}
{Biller}, B.~A., {Males}, J., {Rodigas}, T., {et~al.} 2014, \apjl, 792, L22

\bibitem[{{Boccaletti} {et~al.}(2013){Boccaletti}, {Pantin}, {Lagrange},
  {Augereau}, {Meheut}, \& {Quanz}}]{Boccaletti2013}
{Boccaletti}, A., {Pantin}, E., {Lagrange}, A.-M., {et~al.} 2013, \aap, 560,
  A20

\bibitem[{{Boehler} {et~al.}(2017){Boehler}, {Ricci}, {Weaver}, {Isella},
  {Benisty}, {Carpenter}, {Grady}, {Shen}, {Tang}, \& {Perez}}]{Boehler2017}
{Boehler}, Y., {Ricci}, L., {Weaver}, E., {et~al.} 2017, ArXiv e-prints
  [\eprint[arXiv]{1712.08845}]

\bibitem[{{Bowler} {et~al.}(2014){Bowler}, {Liu}, {Kraus}, \&
  {Mann}}]{Bowler2014}
{Bowler}, B.~P., {Liu}, M.~C., {Kraus}, A.~L., \& {Mann}, A.~W. 2014, \apj,
  784, 65

\bibitem[{{Castelli} \& {Kurucz}(2003)}]{Castelli2003}
{Castelli}, F. \& {Kurucz}, R.~L. 2003, in IAU Symposium, Vol. 210, Modelling
  of Stellar Atmospheres, ed. N.~{Piskunov}, W.~W. {Weiss}, \& D.~F. {Gray},
  A20

\bibitem[{{Chapillon} {et~al.}(2008){Chapillon}, {Guilloteau}, {Dutrey}, \&
  {Pi{\'e}tu}}]{Chapillon2008}
{Chapillon}, E., {Guilloteau}, S., {Dutrey}, A., \& {Pi{\'e}tu}, V. 2008, \aap,
  488, 565

\bibitem[{{Chauvin} {et~al.}(2012){Chauvin}, {Lagrange}, {Beust}, {Bonnefoy},
  {Boccaletti}, {Apai}, {Allard}, {Ehrenreich}, {Girard}, {Mouillet}, \&
  {Rouan}}]{Chauvin2012}
{Chauvin}, G., {Lagrange}, A.-M., {Beust}, H., {et~al.} 2012, \aap, 542, A41

\bibitem[{{Close} {et~al.}(2014){Close}, {Follette}, {Males}, {Puglisi},
  {Xompero}, {Apai}, {Najita}, {Weinberger}, {Morzinski}, {Rodigas}, {Hinz},
  {Bailey}, \& {Briguglio}}]{Close2014}
{Close}, L.~M., {Follette}, K.~B., {Males}, J.~R., {et~al.} 2014, \apjl, 781,
  L30

\bibitem[{{Dong} {et~al.}(2015){Dong}, {Zhu}, {Rafikov}, \& {Stone}}]{Dong2015}
{Dong}, R., {Zhu}, Z., {Rafikov}, R.~R., \& {Stone}, J.~M. 2015, \apjl, 809, L5

\bibitem[{{Fairlamb} {et~al.}(2017){Fairlamb}, {Oudmaijer}, {Mendigutia},
  {Ilee}, \& {van den Ancker}}]{Fairlamb2017}
{Fairlamb}, J.~R., {Oudmaijer}, R.~D., {Mendigutia}, I., {Ilee}, J.~D., \& {van
  den Ancker}, M.~E. 2017, \mnras, 464, 4721

\bibitem[{{Gaia Collaboration} {et~al.}(2016{\natexlab{a}}){Gaia
  Collaboration}, {Brown}, {Vallenari}, {Prusti}, {de Bruijne}, {Mignard},
  {Drimmel}, {Babusiaux}, {Bailer-Jones}, {Bastian}, \& et~al.}]{GaiaDR12016}
{Gaia Collaboration}, {Brown}, A.~G.~A., {Vallenari}, A., {et~al.}
  2016{\natexlab{a}}, \aap, 595, A2

\bibitem[{{Gaia Collaboration} {et~al.}(2016{\natexlab{b}}){Gaia
  Collaboration}, {Prusti}, {de Bruijne}, {Brown}, {Vallenari}, {Babusiaux},
  {Bailer-Jones}, {Bastian}, {Biermann}, {Evans}, \& et~al.}]{Gaia2016}
{Gaia Collaboration}, {Prusti}, T., {de Bruijne}, J.~H.~J., {et~al.}
  2016{\natexlab{b}}, \aap, 595, A1

\bibitem[{{Grady} {et~al.}(2013){Grady}, {Muto}, {Hashimoto}, {Fukagawa},
  {Currie}, {Biller}, {Thalmann}, {Sitko}, {Russell}, {Wisniewski}, {Dong},
  {Kwon}, {Sai}, {Hornbeck}, {Schneider}, {Hines}, {Moro Mart{\'{\i}}n},
  {Feldt}, {Henning}, {Pott}, {Bonnefoy}, {Bouwman}, {Lacour}, {Mueller},
  {Juh{\'a}sz}, {Crida}, {Chauvin}, {Andrews}, {Wilner}, {Kraus}, {Dahm},
  {Robitaille}, {Jang-Condell}, {Abe}, {Akiyama}, {Brandner}, {Brandt},
  {Carson}, {Egner}, {Follette}, {Goto}, {Guyon}, {Hayano}, {Hayashi},
  {Hayashi}, {Hodapp}, {Ishii}, {Iye}, {Janson}, {Kandori}, {Knapp}, {Kudo},
  {Kusakabe}, {Kuzuhara}, {Mayama}, {McElwain}, {Matsuo}, {Miyama}, {Morino},
  {Nishimura}, {Pyo}, {Serabyn}, {Suto}, {Suzuki}, {Takami}, {Takato},
  {Terada}, {Tomono}, {Turner}, {Watanabe}, {Yamada}, {Takami}, {Usuda}, \&
  {Tamura}}]{Grady2013}
{Grady}, C.~A., {Muto}, T., {Hashimoto}, J., {et~al.} 2013, \apj, 762, 48

\bibitem[{{Gullbring} {et~al.}(1998){Gullbring}, {Hartmann}, {Brice{\~n}o}, \&
  {Calvet}}]{Gullbring1998}
{Gullbring}, E., {Hartmann}, L., {Brice{\~n}o}, C., \& {Calvet}, N. 1998, \apj,
  492, 323

\bibitem[{{Hu{\'e}lamo} {et~al.}(2011){Hu{\'e}lamo}, {Lacour}, {Tuthill},
  {Ireland}, {Kraus}, \& {Chauvin}}]{Huelamo2011}
{Hu{\'e}lamo}, N., {Lacour}, S., {Tuthill}, P., {et~al.} 2011, \aap, 528, L7

\bibitem[{{Isella} {et~al.}(2010){Isella}, {Natta}, {Wilner}, {Carpenter}, \&
  {Testi}}]{Isella2010}
{Isella}, A., {Natta}, A., {Wilner}, D., {Carpenter}, J.~M., \& {Testi}, L.
  2010, \apj, 725, 1735

\bibitem[{{Kraus} \& {Ireland}(2012)}]{Kraus2012}
{Kraus}, A.~L. \& {Ireland}, M.~J. 2012, \apj, 745, 5

\bibitem[{{Kurucz}(1993)}]{Kurucz1993}
{Kurucz}, R. 1993, SYNTHE Spectrum Synthesis Programs and Line Data.~Kurucz
  CD-ROM No.~18.~Cambridge, Mass.: Smithsonian Astrophysical Observatory,
  1993., 18

\bibitem[{{Lafreni{\`e}re} {et~al.}(2007){Lafreni{\`e}re}, {Marois}, {Doyon},
  {Nadeau}, \& {Artigau}}]{Lafreniere2007}
{Lafreni{\`e}re}, D., {Marois}, C., {Doyon}, R., {Nadeau}, D., \& {Artigau},
  {\'E}. 2007, \apj, 660, 770

\bibitem[{{Ligi} {et~al.}(2018){Ligi}, {Vigan}, {Gratton}, {de Boer},
  {Benisty}, {Boccaletti}, {Quanz}, {Meyer}, {Ginski}, {Sissa}, {Gry},
  {Henning}, {Beuzit}, {Biller}, {Bonnefoy}, {Chauvin}, {Cheetham}, {Cudel},
  {Delorme}, {Desidera}, {Feldt}, {Galicher}, {Girard}, {Janson}, {Kasper},
  {Kopytova}, {Lagrange}, {Langlois}, {Lecoroller}, {Maire}, {M{\'e}nard},
  {Mesa}, {Peretti}, {Perrot}, {Pinilla}, {Pohl}, {Rouan}, {Stolker},
  {Samland}, {Wahhaj}, {Wildi}, {Zurlo}, {Buey}, {Fantinel}, {Fusco}, {Jaquet},
  {Moulin}, {Ramos}, {Suarez}, \& {Weber}}]{Ligi2018}
{Ligi}, R., {Vigan}, A., {Gratton}, R., {et~al.} 2018, \mnras, 473, 1774

\bibitem[{{Marino} {et~al.}(2015){Marino}, {Casassus}, {Perez}, {Lyra},
  {Roman}, {Avenhaus}, {Wright}, \& {Maddison}}]{Marino2015}
{Marino}, S., {Casassus}, S., {Perez}, S., {et~al.} 2015, \apj, 813, 76

\bibitem[{{Marois} {et~al.}(2006){Marois}, {Lafreni{\`e}re}, {Doyon},
  {Macintosh}, \& {Nadeau}}]{marois2006}
{Marois}, C., {Lafreni{\`e}re}, D., {Doyon}, R., {Macintosh}, B., \& {Nadeau},
  D. 2006, \apj, 641, 556

\bibitem[{{Mawet} {et~al.}(2014){Mawet}, {Milli}, {Wahhaj}, {Pelat}, {Absil},
  {Delacroix}, {Boccaletti}, {Kasper}, {Kenworthy}, {Marois}, {Mennesson}, \&
  {Pueyo}}]{Mawet2014}
{Mawet}, D., {Milli}, J., {Wahhaj}, Z., {et~al.} 2014, \apj, 792, 97

\bibitem[{{Mayama} {et~al.}(2012){Mayama}, {Hashimoto}, {Muto}, {Tsukagoshi},
  {Kusakabe}, {Kuzuhara}, {Takahashi}, {Kudo}, {Dong}, {Fukagawa}, {Takami},
  {Momose}, {Wisniewski}, {Follette}, {Abe}, {Akiyama}, {Brandner}, {Brandt},
  {Carson}, {Egner}, {Feldt}, {Goto}, {Grady}, {Guyon}, {Hayano}, {Hayashi},
  {Hayashi}, {Henning}, {Hodapp}, {Ishii}, {Iye}, {Janson}, {Kandori}, {Kwon},
  {Knapp}, {Matsuo}, {McElwain}, {Miyama}, {Morino}, {Moro-Martin},
  {Nishimura}, {Pyo}, {Serabyn}, {Suto}, {Suzuki}, {Takato}, {Terada},
  {Thalmann}, {Tomono}, {Turner}, {Watanabe}, {Yamada}, {Takami}, {Usuda}, \&
  {Tamura}}]{Mayama2012}
{Mayama}, S., {Hashimoto}, J., {Muto}, T., {et~al.} 2012, \apjl, 760, L26

\bibitem[{{Meeus} {et~al.}(2012){Meeus}, {Montesinos}, {Mendigut{\'{\i}}a},
  {Kamp}, {Thi}, {Eiroa}, {Grady}, {Mathews}, {Sandell}, {Martin-Za{\"i}di},
  {Brittain}, {Dent}, {Howard}, {M{\'e}nard}, {Pinte}, {Roberge},
  {Vandenbussche}, \& {Williams}}]{Meeus2012}
{Meeus}, G., {Montesinos}, B., {Mendigut{\'{\i}}a}, I., {et~al.} 2012, \aap,
  544, A78

\bibitem[{{Mendigut{\'{\i}}a} {et~al.}(2011){Mendigut{\'{\i}}a}, {Calvet},
  {Montesinos}, {Mora}, {Muzerolle}, {Eiroa}, {Oudmaijer}, \&
  {Mer{\'{\i}}n}}]{Mendigutia2011}
{Mendigut{\'{\i}}a}, I., {Calvet}, N., {Montesinos}, B., {et~al.} 2011, \aap,
  535, A99

\bibitem[{{Mendigut{\'{\i}}a} {et~al.}(2017){Mendigut{\'{\i}}a}, {Oudmaijer},
  {Garufi}, {Lumsden}, {Hu{\'e}lamo}, {Cheetham}, {de Wit}, {Norris}, {Olguin},
  \& {Tuthill}}]{Mendigutia2017}
{Mendigut{\'{\i}}a}, I., {Oudmaijer}, R.~D., {Garufi}, A., {et~al.} 2017, \aap,
  608, A104

\bibitem[{{Patat} {et~al.}(2011){Patat}, {Moehler}, {O'Brien}, {Pompei},
  {Bensby}, {Carraro}, {de Ugarte Postigo}, {Fox}, {Gavignaud}, {James},
  {Korhonen}, {Ledoux}, {Randall}, {Sana}, {Smoker}, {Stefl}, \&
  {Szeifert}}]{Patat2011}
{Patat}, F., {Moehler}, S., {O'Brien}, K., {et~al.} 2011, \aap, 527, A91

\bibitem[{{P{\'e}rez} {et~al.}(2016){P{\'e}rez}, {Carpenter}, {Andrews},
  {Ricci}, {Isella}, {Linz}, {Sargent}, {Wilner}, {Henning}, {Deller},
  {Chandler}, {Dullemond}, {Lazio}, {Menten}, {Corder}, {Storm}, {Testi},
  {Tazzari}, {Kwon}, {Calvet}, {Greaves}, {Harris}, \& {Mundy}}]{Perez2016}
{P{\'e}rez}, L.~M., {Carpenter}, J.~M., {Andrews}, S.~M., {et~al.} 2016,
  Science, 353, 1519

\bibitem[{{Pinilla} {et~al.}(2015){Pinilla}, {de Boer}, {Benisty},
  {Juh{\'a}sz}, {de Juan Ovelar}, {Dominik}, {Avenhaus}, {Birnstiel}, {Girard},
  {Huelamo}, {Isella}, \& {Milli}}]{Pinilla2015}
{Pinilla}, P., {de Boer}, J., {Benisty}, M., {et~al.} 2015, \aap, 584, L4

\bibitem[{{Pohl} {et~al.}(2017){Pohl}, {Benisty}, {Pinilla}, {Ginski}, {de
  Boer}, {Avenhaus}, {Henning}, {Zurlo}, {Boccaletti}, {Augereau}, {Birnstiel},
  {Dominik}, {Facchini}, {Fedele}, {Janson}, {Keppler}, {Kral}, {Langlois},
  {Ligi}, {Maire}, {M{\'e}nard}, {Meyer}, {Pinte}, {Quanz}, {Sauvage},
  {Sezestre}, {Stolker}, {Szul{\'a}gyi}, {van Boekel}, {van der Plas},
  {Villenave}, {Baruffolo}, {Baudoz}, {Le Mignant}, {Maurel}, {Ramos}, \&
  {Weber}}]{Pohl2017}
{Pohl}, A., {Benisty}, M., {Pinilla}, P., {et~al.} 2017, \apj, 850, 52

\bibitem[{{Quanz} {et~al.}(2013){Quanz}, {Amara}, {Meyer}, {Kenworthy},
  {Kasper}, \& {Girard}}]{Quanz2013}
{Quanz}, S.~P., {Amara}, A., {Meyer}, M.~R., {et~al.} 2013, \apjl, 766, L1

\bibitem[{{Racine} {et~al.}(1999){Racine}, {Walker}, {Nadeau}, {Doyon}, \&
  {Marois}}]{racine1999}
{Racine}, R., {Walker}, G.~A.~H., {Nadeau}, D., {Doyon}, R., \& {Marois}, C.
  1999, \pasp, 111, 587

\bibitem[{{Reggiani} {et~al.}(2017){Reggiani}, {Christiaens}, {Absil}, {Mawet},
  {Huby}, {Choquet}, {Gomez Gonzalez}, {Ruane}, {Femenia}, {Serabyn},
  {Matthews}, {Barraza}, {Carlomagno}, {Defr{\`e}re}, {Delacroix}, {Habraken},
  {Jolivet}, {Karlsson}, {Orban de Xivry}, {Piron}, {Surdej}, {Vargas Catalan},
  \& {Wertz}}]{Reggiani2017}
{Reggiani}, M., {Christiaens}, V., {Absil}, O., {et~al.} 2017, ArXiv e-prints
  [\eprint[arXiv]{1710.11393}]

\bibitem[{{Rieke} \& {Lebofsky}(1985)}]{Rieke85}
{Rieke}, G.~H. \& {Lebofsky}, M.~J. 1985, \apj, 288, 618

\bibitem[{{Rigliaco} {et~al.}(2012){Rigliaco}, {Natta}, {Testi}, {Randich},
  {Alcal{\`a}}, {Covino}, \& {Stelzer}}]{Rigliaco2012}
{Rigliaco}, E., {Natta}, A., {Testi}, L., {et~al.} 2012, \aap, 548, A56

\bibitem[{{Sallum} {et~al.}(2015){Sallum}, {Follette}, {Eisner}, {Close},
  {Hinz}, {Kratter}, {Males}, {Skemer}, {Macintosh}, {Tuthill}, {Bailey},
  {Defr{\`e}re}, {Morzinski}, {Rodigas}, {Spalding}, {Vaz}, \&
  {Weinberger}}]{Sallum2015}
{Sallum}, S., {Follette}, K.~B., {Eisner}, J.~A., {et~al.} 2015, \nat, 527, 342

\bibitem[{{Schmid} {et~al.}(2017){Schmid}, {Bazzon}, {Milli}, {Roelfsema},
  {Engler}, {Mouillet}, {Lagadec}, {Sissa}, {Sauvage}, {Ginski}, {Baruffolo},
  {Beuzit}, {Boccaletti}, {Bohn}, {Claudi}, {Costille}, {Desidera}, {Dohlen},
  {Dominik}, {Feldt}, {Fusco}, {Gisler}, {Girard}, {Gratton}, {Henning},
  {Hubin}, {Joos}, {Kasper}, {Langlois}, {Pavlov}, {Pragt}, {Puget}, {Quanz},
  {Salasnich}, {Siebenmorgen}, {Stute}, {Suarez}, {Szul{\'a}gyi}, {Thalmann},
  {Turatto}, {Udry}, {Vigan}, \& {Wildi}}]{Schmid2017}
{Schmid}, H.~M., {Bazzon}, A., {Milli}, J., {et~al.} 2017, \aap, 602, A53

\bibitem[{{Soummer} {et~al.}(2012){Soummer}, {Pueyo}, \&
  {Larkin}}]{Soummer2012}
{Soummer}, R., {Pueyo}, L., \& {Larkin}, J. 2012, \apjl, 755, L28

\bibitem[{{Thalmann} {et~al.}(2008){Thalmann}, {Schmid}, {Boccaletti},
  {Mouillet}, {Dohlen}, {Roelfsema}, {Carbillet}, {Gisler}, {Beuzit}, {Feldt},
  {Gratton}, {Joos}, {Keller}, {Kragt}, {Pragt}, {Puget}, {Rigal}, {Snik},
  {Waters}, \& {Wildi}}]{thalmann2008}
{Thalmann}, C., {Schmid}, H.~M., {Boccaletti}, A., {et~al.} 2008, in \procspie,
  Vol. 7014, Ground-based and Airborne Instrumentation for Astronomy II, 70143F

\bibitem[{{van Leeuwen}(2007)}]{vanLeeuwen2007}
{van Leeuwen}, F. 2007, \aap, 474, 653

\bibitem[{{Whelan} {et~al.}(2015){Whelan}, {Hu{\'e}lamo}, {Alcal{\'a}},
  {Lillo-Box}, {Bouy}, {Barrado}, {Bouvier}, \& {Mer{\'{\i}}n}}]{Whelan2015}
{Whelan}, E.~T., {Hu{\'e}lamo}, N., {Alcal{\'a}}, J.~M., {et~al.} 2015, \aap,
  579, A48

\bibitem[{{Zhou} {et~al.}(2014){Zhou}, {Herczeg}, {Kraus}, {Metchev}, \&
  {Cruz}}]{Zhou2014}
{Zhou}, Y., {Herczeg}, G.~J., {Kraus}, A.~L., {Metchev}, S., \& {Cruz}, K.~L.
  2014, \apjl, 783, L17

\bibitem[{{Zhu}(2015)}]{Zhu2015}
{Zhu}, Z. 2015, \apj, 799, 16

\end{thebibliography}

\end{document}